\documentclass[twocolumn,showpacs,showkeys,preprintnumbers]{revtex4-1}
\usepackage{amssymb}

\usepackage{amsfonts}

\usepackage{latexsym}

\usepackage{dcolumn}

\usepackage{amsmath}

\usepackage{graphicx}

\usepackage{psfrag,pstricks}

\begin{document}

\title{The effect of atomic collisions on the quantum phase transition of a Bose-Einstein condensate inside an optical cavity}

\author{A. Dalafi$^{1}$ }
\email{adalafi@yahoo.co.uk}

\author{M. H. Naderi$^{1,2}$}
\author{M. Soltanolkotabi$^{1,2}$}

\affiliation{$^{1}$ Department of Physics, Faculty of Science, University of Isfahan, Hezar Jerib, 81746-73441, Isfahan, Iran\\
$^{2}$Quantum Optics Group, Department of Physics, Faculty of Science, University of Isfahan, Hezar Jerib, 81746-73441, Isfahan, Iran}

\date{\today}

\begin{abstract}
In this paper, we investigate the effect of atomic collisions on the phase transition form the normal to the superradiant phase in a one-dimensional Bose-Einstein condensate (BEC) trapped inside an optical cavity. Specifically, we show that driving the atoms from the side of the cavity leads to the excitation of modes in the edges of the first Brillouin zone of every energy band, which results in the two-mode approximation of the BEC matter field in the limit of weak coupling regime. The nonlinear effect of atom-atom interaction shifts the threshold of the quantum phase transition of the BEC and also affect the power low behavior of quantum fluctuations in the total particle number. Besides, we show the possibility of controlling the quantum phase transition of the system through the \textit{s}-wave scattering frequency when the the strength of the transverse pumping has been fixed.
\end{abstract}

\maketitle

\section{Introduction}
Ultracold atoms trapped inside high-Q cavities are suitable systems for studying the interaction of light with matter in the regime where not only their mutual effects are simultaneously observable but also their quantum mechanical properties are manifested in the same level \citep{Maschler2008,dom JOSA}. Such an interaction is inherently nonlinear which is due to the mutual matter-light coupling \cite{Zhang 2009}. On the other hand, if the density of the atomic ensemble is high enough then another kind of nonlinearity manifests itself which is due to the atom-atom interaction. Such a situation is usually manifested in a BEC \cite{Morsch}. 

Furthermore, ultracold atoms trapped inside optical cavities exhibit phenomena typical of solid-state physics like the formation of energy bands \cite{Bha Opt. Commun} and Bloch oscillations \cite{Prasanna13}. One of the other features of these so-called hybrid systems is their similarities with the optomechanical systems (optical cavities with moving mirror) \cite{Genes 2008, Barzanjeh2}. In the hybrid system the excitation of a collective mode of the cold gas plays the role of the vibrational mode of the moving mirror of the optomechanical cavity \cite{Kanamoto,Brenn Nature,Gupta,Brenn Science,Ritter Appl. Phys. B}. Besides, the nonlinear effects of atom-atom interaction in systems consisting of BEC affect the optical bistability \cite{dalafi 2} and the squeezing of the vibrational modes of the system \cite{dalafi 3}.

\begin{figure}[ht]
\centering
\includegraphics[width=3.8in]{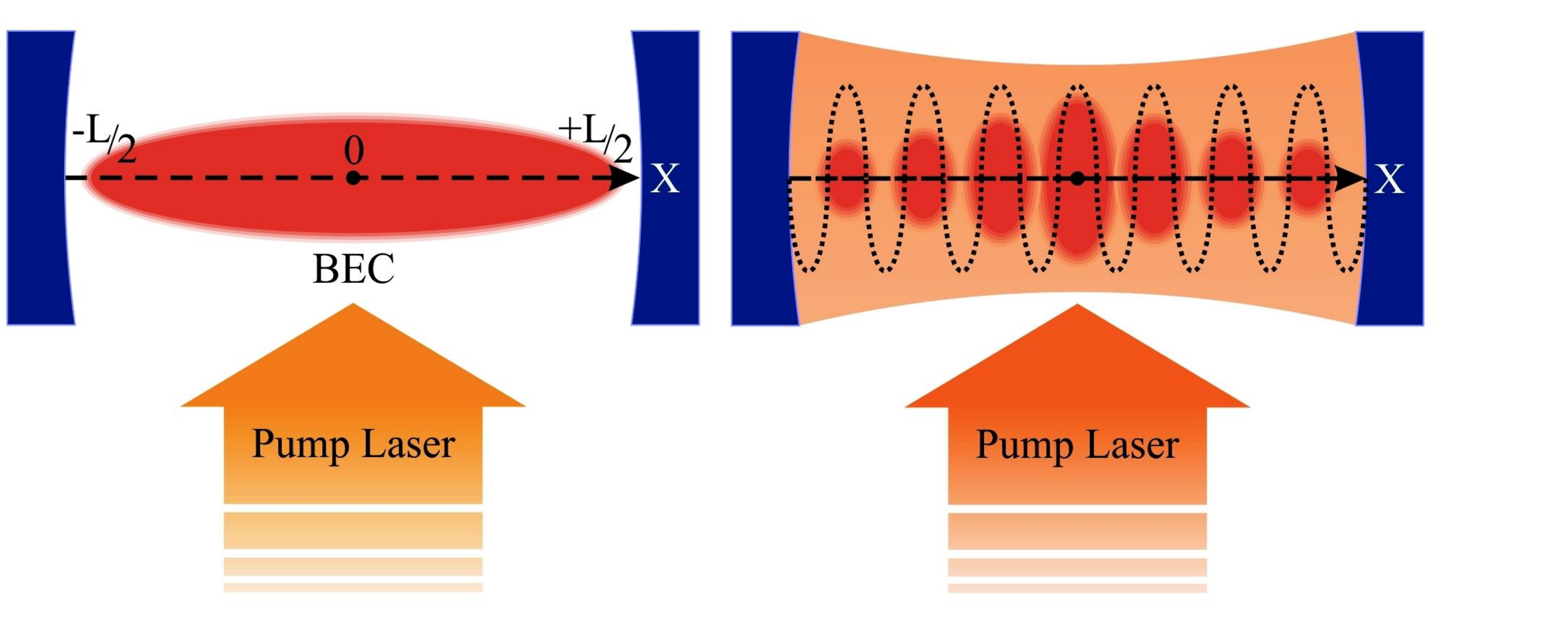}
\caption{
(Color online) The quantum phase transition of a BEC inside an optical cavity from the normal phase (left), when the transverse pumping rate $ \eta_{t} $ is below the critical value $ \eta_{c} $, to the superradiant  phase (right) when $ \eta_{t}> \eta_{c} $.}
\label{fig:fig1}
\end{figure}

Hybrid systems consisting of BEC have also attracted considerable attention in connection with quantum phase transition phenomena. One celebrated example is the transition from a superfluid to a Mott insulator phase in the Bose-Hubbard model \cite{Jaksch98, Bha PRA09,Maschler PRL05, k zhang optcommun} that has been observed experimentally in a gas of ultracold atoms trapped inside an optical lattice \cite{Greiner2002}. Another kind of quantum phase transition from the homogeneous into a periodically patterned distribution can be observed in hybrid systems consisting of a BEC whose atoms are coherently pumped from the side of the cavity [Fig.\ref{fig:fig1}]. It has been shown \cite{nagy PRL10} that this kind of phase transition is very similar to that of the Dicke model from the normal to the superradiant phase. When the intensity of the transverse pump laser is below a critical value, the system is in the normal phase where all the atoms populate the ground state of the optical lattice (distributed homogeneously) and the radiation field inside the cavity is in the vacuum state. However, above the critical point some atoms are excited to higher motional modes and the mean value of the optical field gets nonzero (superradiant phase). In this phase, the spatial distribution of atoms gets a periodic pattern \cite{Konya Domokos 2011}. 

Near the critical point, the second-order correlation functions of the system show a power law behavior. This behavior has been investigated for the stationary state of the driven and damped hybrid system in the thermodynamic limit \cite{nagy PRA11} and also for finite-size systems consisting of low-density Bose-Einstein condensates where the effect of atomic collisions is dispensable \cite{konya PRA12}.

In this paper we are going to study the nonlinear effect of atom-atom interaction on the threshold of quantum phase transition of a BEC trapped inside an optical cavity when the atoms are driven coherently from the side of the cavity. Besides, we show that the atomic collisions can affect the scaling law behavior of quantum fluctuations in the total number of atoms. In addition to the effect of atomic collisions, here we also consider the effect of the damping of the matter field of the BEC due to the leakage of atoms to other modes \cite{k zhang 10}. For this purpose we consider the discrete mode approximation for the BEC like that was studied in our previous paper \cite{dalafi1} with the difference that here we no longer consider the Bogoliubov approximation. 

In Refs.\cite{dalafi1, Szirmai 2010} it has been shown how the atoms are scattered to the modes with momenta $ 2n\hbar k $ when the cavity is pumped from one of its mirrors ($ k $ is the wave number of the optical mode and $ n $ is the band index). Here, using the theoretical description of the band structure of a one-dimensional BEC inside an optical lattice developed in the two above-mentioned references, we will specifically show that driving the atoms from the side of the cavity leads to the excitation of modes in the edges of the first Brillouin zone of every energy band. It is shown that the linear combination of modes in the two opposite edges in the Brillouin zones of two adjacent energy bands forms the modes with momenta $ n\hbar k $ which are coupled to the central modes (with zero quasi mementum) of each energy band.

The paper is structured as follows. In Sec.II we describe the Hamiltonian of the system and show how the trapped atoms are scattered to the modes with momenta $ n\hbar k $ due to the transverse pumping. In Sec. III we use the two-mode approximation for describing the BEC and solve for the mean-field values of the system through the numerical self-consistent method considering atom-atom interaction. Then, we investigate the effect of atomic collisions on the threshold of quantum phase transition and on the power law behavior of quantum fluctuations. Furthermore, we investigate the quantum phase transition of the system in terms of the \textit{s}-wave scatterin frequency when the strength of the transverse pumping has been kept on a fixed value. Finally, our conclusions are summarized in Sec. IV.

\section{Description of the system}\label{secH}
We consider a system consisting of a BEC of $N$ two-level atoms inside an optical cavity with length $L$ where the atoms are coherently driven from the side by a laser with frequency $\omega_{p}$, and wave number $k=\omega_{p}/c$, directed perpendicularly to the cavity axis (Fig.\ref{fig:fig1}). We assume the BEC to be confined in a cylindrically symmetric trap with a transverse trapping frequency $\omega_{\mathrm{\perp}}$ and negligible longitudinal confinement along the $x$ direction \cite{Morsch}. In this way we can describe the dynamics within an effective one-dimensional model by quantizing the atomic motional degree of freedom along the $x$ axis only.

In the dispersive regime, where the laser pump is detuned far below the atomic resonance (the absolute value of $\Delta_{a}=\omega_{p}-\omega_{a}<0$  exceeds the atomic linewidth $\gamma$ by orders of magnitude), the excited electronic state of the atoms can be adiabatically eliminated and spontaneous emission can be neglected \cite{Masch Ritch 2004}. In the frame rotating at the pump frequency, the many-body Hamiltonian reads
\begin{eqnarray}\label{H1}
&H&=-\hbar\Delta_{c} a^{\dagger} a+\int_{-\frac{L}{2}}^{\frac{L}{2}}\Psi^{\dagger}(x)\Big[\frac{-\hbar^{2}}{2m_{a}}\frac{d^{2}}{dx^{2}}+\hbar U_{0} \cos^2(kx) a^{\dagger} a\nonumber\\
&&+\hbar\eta_{t}\cos(kx)(a^{\dagger}+a)+\frac{1}{2} U_{s}\Psi^{\dagger}(x)\Psi(x)\Big] \Psi(x) dx.
\end{eqnarray}
Here, $a$ is the annihilation operator of the optical field, $\Delta_{c}=\omega_{p}-\omega_{c}$ is the cavity-pump detuning, $U_{0}=g_{0}^{2}/\Delta_{a}$ is the optical lattice barrier height per photon which represents the atomic back action on the field, $g_{0}$ is the vacuum Rabi frequency, $U_{s}=\frac{4\pi\hbar^{2} a_{s}}{m_{a}w^{2}}$, $a_{s}$ is the two-body \textit{s}-wave scattering length \cite{Masch Ritch 2004,Dom JB}, and $ w $ is the waist of the optical potential.The first term in the second line of Eq.(\ref{H1}) describes the effect of the transverse pump field which drives the atoms with the constant amplitude $ \eta_{t}=\Omega_{R}g/\Delta_{a} $ where $ \Omega_{R} $ is the Rabi frequency.

\subsection{The matter field of the BEC}
As is shown in Ref.\cite{dalafi1} the matter field of the BEC can be expanded in terms of plane waves in the following way
\begin{equation}\label{Psib}
\Psi(x)=\frac{1}{\sqrt{L}}\sum_{n\in \mathbb{Z}}\sum_{m=-l/2}^{+l/2} b_{n,m} e^{i(n+m/l)2kx},
\end{equation}
where $ l=2L/\lambda $ is the number of periods of the matter field inside the cavity which is assumed to be even, and $ b_{n,m} $ is the annihilation operator for the atomic field that annihilates a particle in a state determined with the Bloch band index $ n $ and quasimomentum $ q_{m}=2m k/l $.

In the limit of weak photon-atom coupling, where $U_{0}\langle a^{\dagger}a\rangle\leq 10\omega_{R}$ ($\omega_{R}=\frac{\hslash k^{2}}{2m_{a}}$ is the recoil frequency of the condensate atoms), the above expansion can be restricted to the lowest band numbers $ n=0,\pm1 $. Besides, if the system starts from a homogeneous BEC, only the cosine parts of the exponential functions are excited because of the parity conservation\cite{Nagy Ritsch 2009}. Therefore, the matter field can be written in the following form 
\begin{eqnarray}\label{PsiC}
&\Psi(x)&=\sqrt{\frac{1}{L}}C_{00}+\sqrt{\frac{2}{L}}C_{10}\cos(2kx)\nonumber\\
&&+\sqrt{\frac{2}{L}}\sum_{m=1}^{l/2}\Big[C_{0m}\cos\Big(\frac{m}{l}2kx\Big)\nonumber\\
&&+C_{1m}\cos\Big(1+\frac{m}{l}\Big)2kx+C_{1,-m}\cos\Big(1-\frac{m}{l}\Big)2kx\Big],\nonumber\\
\end{eqnarray}
where the new modes $ C_{nm} $ have been defined in terms of the following Bogoliubov transformations
\begin{eqnarray}\label{Cnm}
C_{00}&=&b_{00},\nonumber\\
C_{0m}&=&\frac{1}{\sqrt{2}}(b_{0m}+b_{0,-m})=C_{0,-m},\nonumber\\
C_{1m}&=&\frac{1}{\sqrt{2}}(b_{1m}+b_{-1,-m}),\nonumber\\
C_{1,-m}&=&\frac{1}{\sqrt{2}}(b_{1,-m}+b_{-1,m}).
\end{eqnarray}
Confining the matter field to some discrete modes like Eq.(\ref{PsiC}), is called the discrete-mode approximation (DMA) \cite{Zhang 2009}. In the following we will derive the Hamiltonian (\ref{H1}) in terms of the discrete modes.

\subsection{System Hamiltonian in the discrete-mode approximation}
In order to see which modes of the BEC are excited due to the transverse pumping of the atoms let us first investigate that part of Eq.(\ref{H1}) which is related to the transverse pumping i.e.,
\begin{equation}\label{Htp0}
H_{tp}=\hbar\eta_{t}(a^{\dagger}+a)\int_{-L/2}^{L/2}\Psi^{\dagger}(x)\cos(kx)\Psi(x) dx.
\end{equation}
Substituting Eq.(\ref{PsiC}) into Eq.(\ref{Htp0}) we will have
\begin{eqnarray}\label{HtpC}
&H_{tp}&=\hbar\eta_{t} (a+a^{\dagger})\times\nonumber\\
&&\Big[\frac{\sqrt{2}}{2}(C^{\dagger}_{0,l/2}+C^{\dagger}_{1,-l/2})C_{00}+\frac{1}{2}(C^{\dagger}_{0,l/2}+C^{\dagger}_{1,-l/2})C_{10}\nonumber\\
&&+\frac{\sqrt{2}}{2}C^{\dagger}_{00}(C_{0,l/2}+C_{1,-l/2})+\frac{1}{2}C^{\dagger}_{10}(C_{0,l/2}+C_{1,-l/2})\Big].\nonumber\\
\end{eqnarray}

As is seen from this equation, driving the atoms from the side of the cavity makes the modes $ C_{10}, C_{0,l/2} $ and $ C_{1,-l/2} $ be excited while leaving the other modes unexcited. Besides, it leads to a coupling between the linear combination $ C_{0,l/2}+C_{1,-l/2} $ and the two modes $ C_{00} $ and $ C_{10} $ (the central modes of the bands $ n=0 $ and $ n=1 $). Based on Eq.(\ref{PsiC}) both $ C_{0,l/2} $ (the mode in the tail of the Brillouin zone of the band $ n=0 $) and $ C_{1,-l/2} $ (the mode in the head of the Brillouin zone of the band $ n=1 $) have the same mode function, i.e, $ \cos(kx) $. Therefore, if we consider $ c_{1}\equiv C_{0,l/2}+C_{1,-l/2}$ as a new mode with quasimomentum $ \hbar k $ and redefine the two central modes with momenta $ 0 $ and $ 2\hbar k $ by $ c_{0}\equiv C_{00} $ and $ c_{2}\equiv C_{10} $ respectively, then Eq.(\ref{HtpC}) reads
\begin{equation}\label{Htpc}
H_{tp}=\frac{1}{2}\hbar\eta_{t} (a+a^{\dagger})\Big[\sqrt{2}(c^{\dagger}_{1}c_{0}+c^{\dagger}_{0}c_{1})+(c^{\dagger}_{1}c_{2}+c^{\dagger}_{2}c_{1})\Big].
\end{equation}
Considering higher bands of energy, it can be easily seen that other modes on the edges of the Brillouin zone defined as $ c_{2n+1}\equiv C_{n,l/2}+C_{n+1,-l/2} $ are coupled to the central modes $ c_{2n}\equiv C_{n,0} $ for $ (n=1,2,...) $ which leads to the appearnce of terms as $ c^{\dagger}_{n}c_{n+1}+c^{\dagger}_{n+1}c_{n} $ inside the brackets of Eq.(\ref{Htpc}). Therefore, by pumping the atoms transversely the modes in the edges of the first Brillouin zones of the energy bands are excited so that the linear combination of modes in the two opposite edges in the Brillouin zones of two adjacent energy bands form the modes with momenta $ n\hbar k $ which are coupled to the central modes (with zero quasi mementum) of each energy band.

In the same way one can calculate that part of the Hamiltonian (\ref{H1}) corresponding to the interaction between the atoms and intracavity photons, i.e,.
\begin{equation}\label{Hatph0}
H_{at-ph}=\hbar U_{0}a^{\dagger}a \int_{-L/2}^{L/2}\Psi^{\dagger}(x)\cos^{2}(kx)\Psi(x) dx.
\end{equation}
Again, substituting Eq.(\ref{PsiC}) into Eq.(\ref{Hatph0}) and after some simplification we will have
\begin{eqnarray}\label{Hatphc}
H_{at-ph}&=&\frac{1}{2}\hbar U_{0}a^{\dagger}a\times\nonumber\\
&&\Big[c^{\dagger}_{0}c_{0}+\frac{3}{2}c^{\dagger}_{1}c_{1}+c^{\dagger}_{2}c_{2}+\frac{1}{\sqrt{2}}(c^{\dagger}_{0}c_{2}+c^{\dagger}_{2}c_{0})+B\Big],\nonumber\\
\end{eqnarray}
in which the operator $ B $ has been defined as
\begin{eqnarray}\label{B}
&B&=\sum_{m=1}^{l/2-1}\Big[C^{\dagger}_{0m}C_{0m}+C^{\dagger}_{1m}C_{1m}+C^{\dagger}_{1,-m}C_{1,-m}\nonumber\\
&&+\frac{1}{2}(C^{\dagger}_{0m}C_{1m}+C^{\dagger}_{0m}C_{1,-m}+C^{\dagger}_{1,m}C_{0m}+C^{\dagger}_{1,-m}C_{0m})\Big].\nonumber\\
\end{eqnarray}

Finally, the kinetic energy part of the Hamiltonian (the first term inside the integral in Eq.(\ref{H1})) takes the following form
\begin{equation}\label{Hkin}
H_{kin}=\hbar\omega_{R}(c^{\dagger}_{1}c_{1}+4c^{\dagger}_{2}c_{2}+4F),
\end{equation}
where
\begin{eqnarray}
F&=&\sum_{m=1}^{l/2-1}\Big[\Big(\frac{m}{l}\Big)^{2}C^{\dagger}_{0m}C_{0m}+\Big(1+\frac{m}{l}\Big)^{2}C^{\dagger}_{1m}C_{1m}\nonumber\\
&&+\Big(1-\frac{m}{l}\Big)^{2}C^{\dagger}_{1,-m}C_{1,-m}\Big].
\end{eqnarray}

Based on Eqs.(\ref{Htpc}),(\ref{Hatphc}), and (\ref{Hkin}) the total Hamiltonian of the system, i.e., Eq.(1), can written as
\begin{eqnarray}\label{Hf}
H&=&-\hbar\Delta_{c} a^{\dagger} a+\hbar\omega_{R}(c^{\dagger}_{1}c_{1}+4c^{\dagger}_{2}c_{2})\nonumber\\
&&+\frac{1}{2}\hbar U_{0}a^{\dagger}a\Big[c^{\dagger}_{0}c_{0}+\frac{3}{2}c^{\dagger}_{1}c_{1}+c^{\dagger}_{2}c_{2}+\frac{1}{\sqrt{2}}(c^{\dagger}_{0}c_{2}+c^{\dagger}_{2}c_{0})\Big]\nonumber\\
&&+\frac{1}{2}\hbar\eta_{t} (a+a^{\dagger})\Big[\sqrt{2}(c^{\dagger}_{1}c_{0}+c^{\dagger}_{0}c_{1})+(c^{\dagger}_{1}c_{2}+c^{\dagger}_{2}c_{1})\Big]\nonumber\\
&&+4\hbar\omega_{R}F+\frac{1}{2}\hbar U_{0}a^{\dagger}a B+H_{aa},
\end{eqnarray}
where $ H_{aa} $ is the atom-atom interaction part of the Hamiltonian which is given by the last term of Eq.(\ref{H1}). The Hamiltonian obtained in Eq.(\ref{Hf}), except for the last three terms, is just like Eq.(3.3) of Ref.\cite{Konya Domokos 2011} and also Eq.(8) of Ref.\cite{Zubairy} .

\subsection{The Heisenberg equations of motion}
Since the photon-field operator $ a $ and the matter-field operators $ c_{0}, c_{1} $ and $ c_{2} $ commute with the operators $ F $ and $ B $, the Heisenberg equations of the matter fields do not change with or without the presence of the terms consisting of these operators in the total Hamiltonian (\ref{Hf}). The only modification due to these additional terms is the term $ -\frac{1}{2}U_{0}aB $  which appears in the Heisenberg equation of the photon-field operator
\begin{eqnarray}\label{Heizenberg a}
\dot{a}&=&i\Big[\delta_{c}+i\kappa-\frac{1}{4}U_{0}(\sqrt{2}(c^{\dagger}_{0}c_{2}+c^{\dagger}_{2}c_{0})+c^{\dagger}_{1}c_{1}+2B)\Big]a\nonumber\\
&&-\frac{1}{2}i\eta_{t}\Big[\sqrt{2}(c^{\dagger}_{1}c_{0}+c^{\dagger}_{0}c_{1})+(c^{\dagger}_{1}c_{2}+c^{\dagger}_{2}c_{1})\Big]+\xi.
\end{eqnarray}
Here, $ \delta_{c}=\Delta_{c}-\frac{1}{2}NU_{0} $ and $ \kappa $ is  the dissipation rate of the cavity field. The cavity-field quantum vacuum fluctuation $\xi(t)$ satisfies the Markovian correlation functions, i.e., $\langle\xi(t)\xi^{\dagger}(t)\rangle=2\kappa(n_{ph}+1)\delta(t-t^{\prime})$, $\langle\xi^{\dagger}(t^{\prime})\xi(t))\rangle=2\kappa n_{ph}\delta(t-t^{\prime})$ with the average thermal photon number $n_{ph}$ which is nearly zero at optical frequencies \cite{Gardiner}. As is seen From Eq.(\ref{Heizenberg a}), the optical mode is coupled to the modes $ C_{0,m} $ and $ C_{1,m} $ through the operator $ B $. On the other hand, the Heisenberg equations for $ C_{0,m} $, $ C_{1,m} $ without considering the last term of the Hamiltonian (\ref{Hf}), i.e, $ H_{aa} $, reads
\begin{subequations}\label{Heis C0m}
\begin{eqnarray}
\dot{C}_{0,m}&=&-i4\omega_{R}\Big(\frac{m}{l}\Big)^2 C_{0,m}\nonumber\\
&&-\frac{1}{2}iU_{0}a^{\dagger}a\Big(C_{0,m}+\frac{1}{2}C_{1,m}+\frac{1}{2}C_{1,-m}\Big),\\
\dot{C}_{1,m}&=&-i4\omega_{R}\Big(1+\frac{m}{l}\Big)^2 C_{1,m}\nonumber\\
&&-\frac{1}{2}iU_{0}a^{\dagger}a\Big(C_{1,m}+\frac{1}{2}C_{0,m}\Big).
\end{eqnarray}
\end{subequations}
Based on these equations, there is no direct coupling between the modes $ C_{nm} $ and $ c_{n} $ in the absence of atomic collisions. In fact, the only way that these modes can be coupled to each other in the absence of atom-atom interaction is through the optical mode, i.e., through the term $ \frac{1}{2}\hbar U_{0}a^{\dagger}aB $ in the Hamiltonian (\ref{Hf}), which is negligible in the limit of weak photon-atom coupling. However, due to the atomic interactions the modes $ c_{n} $ can be coupled to the modes $ C_{n,m} $ in a very complicated way. In fact, the modes $ C_{n,m} $ as well as the harmonic trapping potential of the BEC act as a kind of reservoir which lead to damping and broadening of these modes \cite{dalafi1, k zhang  10}. In the rest of this paper we will set aside the modes $ C_{n,m} $ and consider a two-mode approximation for the matter field of the BEC. The effects of the extra modes $ C_{n.m} $ as well as that of the harmonic trapping potential are simulated as a damping process which leads to the fluctuation and dissipation of the two modes of the matter-field.

\section{The two-mode model of the BEC}
Using the approximations mentioned in the previous section one can set aside the terms containing $ F $ and $ B $ in Eq.(\ref{Hf}) and also limit the matter field of the BEC just to the modes $ c_{0} $ and $ c_{1} $ so that it can be written as
\begin{equation}\label{Psi 2mode}
\Psi(x)=\frac{1}{\sqrt{L}}c_{0}+\sqrt{\frac{2}{L}}c_{1}\cos(kx).
\end{equation}
In this way, we will obtain the following Hamiltonian for a two-mode BEC
\begin{eqnarray}\label{H 2mode}
H&=&-\hbar\Delta_{c} a^{\dagger} a+\hbar\omega_{R}c^{\dagger}_{1}c_{1}+\frac{1}{2}\hbar U_{0}a^{\dagger}a\Big(c^{\dagger}_{0}c_{0}+\frac{3}{2}c^{\dagger}_{1}c_{1}\Big)\nonumber\\
&&+\frac{1}{\sqrt{2}}\hbar\eta_{t} (a+a^{\dagger})(c^{\dagger}_{1}c_{0}+c^{\dagger}_{0}c_{1})+\frac{\hbar\omega_{sw}}{4N}\Big(c^{\dagger2}_{0}c^{2}_{0}\nonumber\\
&&+c^{\dagger2}_{0}c^{2}_{1}+c^{\dagger2}_{1}c^{2}_{0}+4c^{\dagger}_{0}c_{0}c^{\dagger}_{1}c_{1}+\frac{3}{2}c^{\dagger2}_{1}c^{2}_{1}\Big).
\end{eqnarray}
Here, the terms proportional to the s-wave scattering frequency, i.e., $ \omega_{sw}=8\pi\hbar a_{s}N/m_{a}L w^{2}$, have come through the atom-atom interaction part, i.e., the last term of Eq.(\ref{H1}). The Heisenberg equations of motion in the grand canonical formalism are written in terms of the grand canonical Hamiltonian, i.e., $ H-\mu\hat{N} $, where $ \mu $ is the chemical potential and
\begin{equation}\label{N}
\hat{N}=\int\Psi^{\dagger}(x)\Psi(x)dx= c^{\dagger}_{0}c_{0}+c^{\dagger}_{1}c_{1}
\end{equation}
is the particle number operator whose expectation value is the mean value of total perticles, i.e., $ \langle\hat{N}\rangle=N $. Therefore, the time evolution of the optical and the matter-field modes are obtained as follows

\begin{subequations}\label{Heis a,c}
\begin{eqnarray}
\dot{a}&=&i\Delta_{c}a-\frac{iU_{0}}{2}a\Big(c^{\dagger}_{0}c_{0}+\frac{3}{2}c^{\dagger}_{1}c_{1}\Big)\nonumber\\
&&-\frac{i}{\sqrt{2}}\eta_{t}(c^{\dagger}_{1}c_{0}+c^{\dagger}_{0}c_{1})-\kappa a+\xi(t),\label{Hza}\\
\dot{c}_{0}&=&\Big[i\Big(\frac{\mu}{\hbar}-\frac{U_{0}}{2}a^{\dagger}a\Big)-\gamma_{c}\Big] c_{0}-\frac{i}{\sqrt{2}}\eta_{t}(a^{\dagger}+a) c_{1}\nonumber\\
&&-\frac{i\omega_{sw}}{2N}\Big(c^{\dagger}_{0}c^{2}_{0}+c^{\dagger}_{0}c^{2}_{1}+2c_{0}c^{\dagger}_{1}c_{1}\Big)+f_{0}(t),\label{Hzc0}\\
\dot{c}_{1}&=&\Big[i\Big(\frac{\mu}{\hbar}-\omega_{R}-\frac{3U_{0}}{4}a^{\dagger}a\Big)-\gamma_{c}\Big] c_{1}-\frac{i}{\sqrt{2}}\eta_{t}(a^{\dagger}+a) c_{0}\nonumber\\
&&-\frac{i\omega_{sw}}{2N}\Big(c^{\dagger}_{1}c^{2}_{0}+2c^{\dagger}_{0}c_{0}c_{1}+\frac{3}{2}c^{\dagger}_{1}c^{2}_{1}\Big)+f_{1}(t),\label{Hzc1}
\end{eqnarray}
\end{subequations}
where, $ \gamma_{c} $ is the dissipation of the collective density excitations of the matter field and $f_{0}(t)$ and $f_{1}(t)$ are the thermal noise inputs for the two modes of BEC which satisfy the same Markovian correlation functions as those of the optical noise, i.e., $\langle f_{j}(t)f_{j}^{\dagger}(t)\rangle=2\gamma_{c}\delta(t-t^{\prime})$, $\langle f_{j}^{\dagger}(t^{\prime})f_{j}(t))\rangle=0$ for $ (j=0,1) $. The noise sources are assumed uncorrelated for the different modes of both the matter and light fields. Now, we can linearize Eqs.(\ref{Hza},\ref{Hzc0},\ref{Hzc1}) by separating the mean values of the operators from the quantum fluctuations. For this purpose, if we substitute $ a=\sqrt{N}\alpha+\delta a, c_{0}=\sqrt{N}\beta_{0}+\delta c_{0} $ and $ c_{1}=\sqrt{N}\beta_{1}+\delta c_{1} $ into Eqs.(\ref{Heis a,c}) we will obtain a set of nonlinear algebraic equations for the steady-state mean values
\begin{subequations}\label{mean}
\begin{eqnarray}
\Big[i(\delta_{c}-u\beta^{2}_{1})-\kappa\Big]\alpha&=&iy\beta_{0}\beta_{1},\label{alpha}\\
\mathbf{M}(\alpha,\beta)\beta&=&\frac{\mu}{\hbar}\beta,\label{beta}
\end{eqnarray}
\end{subequations}
where the parameters $ u=\frac{1}{4}NU_{0}, y=\sqrt{2N}\eta_{t} $, and $ \beta=(\beta_{0},\beta_{1})^{T} $ is a column vector. $ \mathbf{M}(\alpha,\beta) $ is a matrix which reads
\begin{equation}\label{M}
\mathbf{M} =\left(\begin{array}{cc}
2u|\alpha|^2+w_{1} & y\alpha_{R}\\
y\alpha_{R} & \omega_{R}+3u|\alpha|^2+w_{2} \\
\end{array}\right),
\end{equation}
where $ w_{1}=\frac{1}{2}\omega_{sw}(1+2\beta^{2}_{1}) $, $ w_{2}= \frac{3}{2}\omega_{sw}(1-\frac{1}{2}\beta^{2}_{1})$, and $ \alpha_{R} $ is the real part of the complex number $ \alpha $.

On the other hand, the linearized quantum Langevine equations (QLEs) for the quadrature fluctuations of the optical field, i.e., $ \delta X_{a}=\frac{1}{\sqrt{2}}(\delta a+\delta a^{\dagger}), \delta P_{a}=\frac{1}{\sqrt{2}i}(\delta a-\delta a^{\dagger}) $, and the quadrature fluctuations of the matter field, i.e., $ \delta X_{j}=\frac{1}{\sqrt{2}}(\delta c_{j}+\delta c^{\dagger}_{j}), \delta P_{j}=\frac{1}{\sqrt{2}i}(\delta c_{j}-\delta c^{\dagger}_{j}) $ with $ j=(0,1) $, can be written in following compact form:
\begin{equation}\label{deltaZ}
\delta\dot{Z}(t)=A \delta Z(t) +n(t),
\end{equation}
where $ \delta Z=[\delta X_{a},\delta P_{a},\delta X_{0},\delta P_{0},\delta X_{1},\delta P_{1}]^{T} $ is the vector of continuous variable fluctuation operators and
\begin{equation}
n(t)=[\xi_{x}(t),\xi_{p}(t),f_{x0}(t), f_{p0}(t),f_{x1}(t), f_{p1}(t)]^{T},
\end{equation}
is the corresponding vector of noises. Besides, the $ 6\times 6 $ matrix $ A $ is the drift matrix given by
\begin{widetext}
\begin{equation}\label{A}
A=\left(\begin{array}{cccccc}
-\kappa & -\tilde{\delta}_{c} & 4u\beta_{0}\alpha_{I} & 0 & 6u\beta_{1}\alpha_{I} & 0 \\
  \tilde{\delta}_{c} & -\kappa & -(y\beta_{1}+4u\beta_{0}\alpha_{R}) & 0 & -(y\beta_{0}+6u\beta_{1}\alpha_{R}) & 0 \\
    0 & 0 & -\gamma_{c} & \Omega_{0}+\frac{1}{2}\omega_{sw} & 0 & y\alpha_{R}+\beta_{0}\beta_{1}\omega_{sw} \\
   -(y\beta_{1}+4u\beta_{0}\alpha_{R}) & -4u\beta_{0}\alpha_{I} & -(\Omega_{0}+\frac{3}{2}\omega_{sw}) & -\gamma_{c} & -(y\alpha_{R}+3\beta_{0}\beta_{1}\omega_{sw}) & 0\\
  0& 0& 0 & y\alpha_{R}+\beta_{0}\beta_{1}\omega_{sw} &-\gamma_{c} & \Omega_{1} \\
   -(y\beta_{0}+6u\beta_{1}\alpha_{R}) & -6u\beta_{1}\alpha_{I} & -(y\alpha_{R}+3\beta_{0}\beta_{1}\omega_{sw}) & 0 & -\Omega_{1} &-\gamma_{c}
  \end{array}\right),
\end{equation}
\end{widetext}
where
\begin{subequations}
\begin{eqnarray}
\tilde{\delta_{c}}&=&\delta_{c}-u\beta_{1}^{2},\\
\Omega_{0}&=&2u|\alpha|^{2}-\frac{\mu}{\hbar},\\
\Omega_{1}&=&\omega_{R}-\frac{\mu}{\hbar}+3u|\alpha|^{2}+\frac{1}{2}\omega_{sw}\Big(\frac{1}{2}\beta_{1}^{2}+\beta_{1}+3\Big).\nonumber\\
\end{eqnarray}
\end{subequations}

\subsection{Mean field solutions}
Equation (\ref{beta}) is a nonlinear eigenvalue problem in which $ \beta $ is the eigenvector of the matrix $ \mathbf{M}(\alpha,\beta) $ and $ \mu/\hbar $ is the smallest eigenvalue. There are two kinds of nonlinearity in this eigenvalue problem: the first one is due to the indirect dependence of  the matrix $ \mathbf{M} $ on $ \beta $ through the mean optical field $ \alpha $ which is itself dependent on $ \beta $  through Eq.(\ref{alpha}), and the second one which is due to the direct dependence of $ \mathbf{M} $ on $ \beta $ through the parameters $ w_{1} $ and $ w_{2} $, has been originated from the effect of atomic collisions.

\begin{figure}[ht]
\centering
\includegraphics[width=3in]{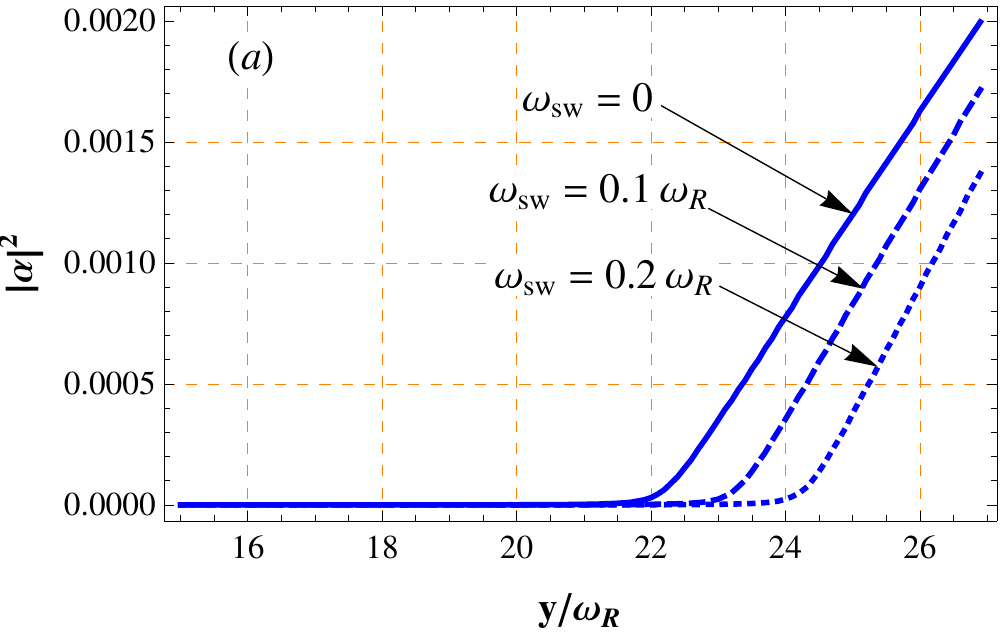}
\includegraphics[width=2.9in]{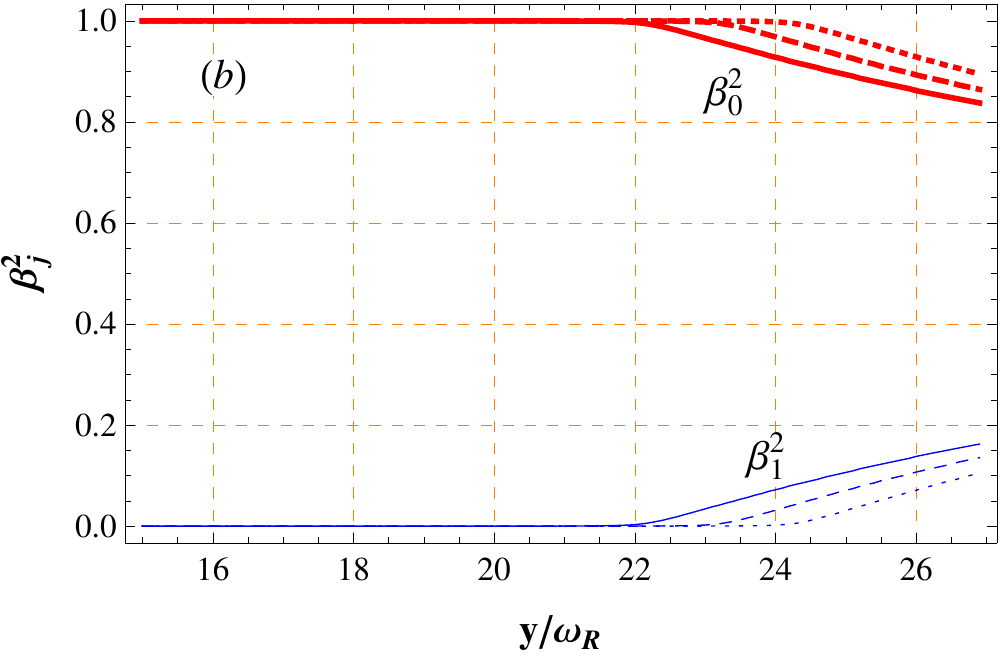}
\caption{
(Color online) (a) The mean photon number of the cavity normalized to $ \sqrt{N} $ and (b) the mean value fraction of atoms in the condensate mode $ c_{0} $ (thick red lines) and the Bogoliubov mode $ c_{1} $ (blue thin lines) versus the pumping strength $ y/\omega_{R} $  for three different values of the \textit{s}-wave scattering frequency: $ \omega_{sw}=0 $ (solid lines), $ \omega_{sw}=0.1\omega_{R} $ (dashed lines), and $ \omega_{sw}=0.2\omega_{R} $ (dotted lines). The parameters are $ \delta_{c}=-100\omega_{R}, u=-20\omega_{R} $, and $ \kappa=200\omega_{R} $.}
\label{fig:fig2}
\end{figure}

\begin{figure}[ht]
\centering
\includegraphics[width=2.8in]{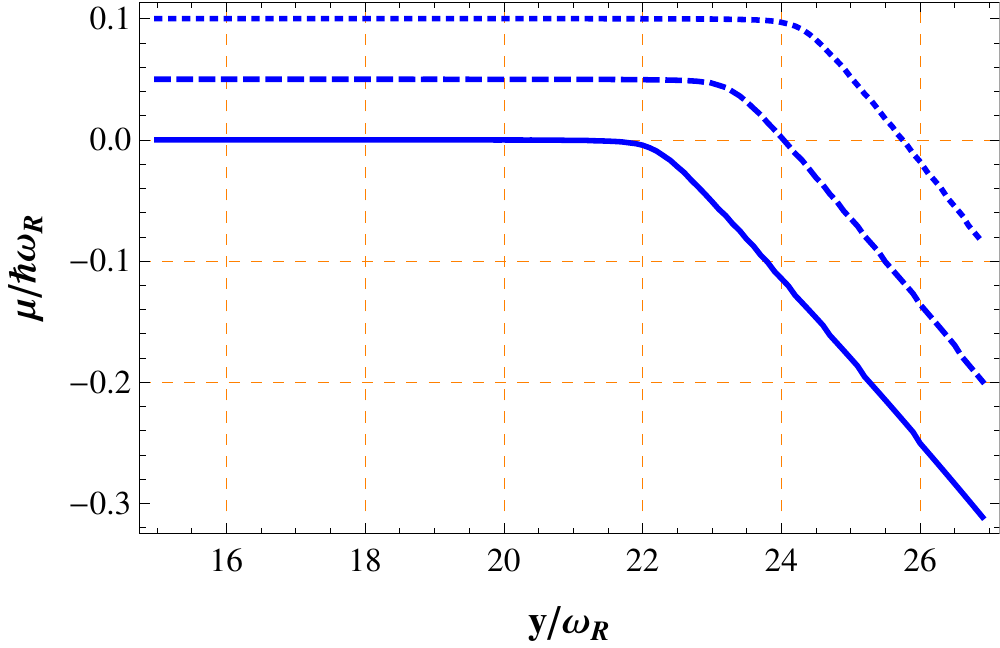}
\caption{
(Color online) The chemical potential of the BEC normalized to $ \hbar\omega_{R} $ versus the pumping strength $ y/\omega_{R} $  for three different values of the \textit{s}-wave scattering frequency: $ \omega_{sw}=0 $ (solid line), $ \omega_{sw}=0.1\omega_{R} $ (dashed line), and $ \omega_{sw}=0.2\omega_{R} $ (dotted line). The parameters are the same as those of Fig.\ref{fig:fig2}.}
\label{fig:fig3}
\end{figure}

Eqs.(\ref{alpha}, \ref{beta}) can be solved by the self-consistent method. For this purpose, we first take arbitrary values for $ \alpha_{R}, \alpha_{I} $, and $ \beta_{1} $ and calculate the smallest eigenvalue of the matrix $ \mathbf{M} $ and obtain its corresponding eigenvector $ \beta $. Next, we substitute the obtained values of $ \beta_{0}, \beta_{1} $ into Eq.(\ref{alpha}) and get the output values for $ \alpha_{R}, \alpha_{I} $. If the difference between the input and the output values of $ \alpha_{R} $ is larger than a specified value $ \epsilon $, we repeat this procedure until this difference becomes less than $ \epsilon $. Whenever this condition is satisfied we stop the procedure and take the last obtained values as the self-consistent solutions. The numerical results of the self-consistent solution to the nonlinear eigenvalue problem [Eqs.(\ref{alpha}, \ref{beta})] have been demonstrated in Fig.\ref{fig:fig2} and Fig.\ref{fig:fig3} in which the value of $ \epsilon $ has be chosen as small as $ 10^{-4} $.

The behavior of quantum phase transition can be seen in Figs.\ref{fig:fig2}(a) and \ref{fig:fig2}(b) which show repectively, $ |\alpha|^2 $, the mean photon number of the cavity normalized to $ \sqrt{N} $,  and the mean value fraction of atoms versus the pumping strength $ y/\omega_{R} $ for three different values of the \textit{s}-wave scattering frequency. As is seen from Fig.\ref{fig:fig2}(a), the mean value of the optical field is zero below a critical value $ y_{c} $ and nonzero above that value. In Fig.\ref{fig:fig2}(b) the mean value fraction of atoms in the condensate mode $ c_{0} $ and the Bogoliubov mode $ c_{1} $ have been plotted versus the pumping strength. As is seen, below the threshold value $ y_{c} $ all the atoms are in the condensate state mode and there is no excitation, but above the critical point the Bogoliubov mode is populated and the quantum phase transition occurs. The results obtained here are in good accordance with those obtained in Ref.\cite{Konya Domokos 2011} in the absence of atom-atom interaction.

The important result is that the value of $ y_{c} $ depends on the strength of atomic interaction, i.e., $ \omega_{sw} $. In the absence of atomic interaction, i.e., for $ \omega_{sw}=0 $, $ y_{c}\simeq22\omega_{R} $ (solid lines), while this value is increased to more than $ 23\omega_{R} $ for $ \omega_{sw}=0.1\omega_{R} $ (dashed lines), and $ 24\omega_{R} $ for $ \omega_{sw}=0.2\omega_{R} $ (dotted lines). In this way, the threshold of the phase transition of the system can be controlled by the \textit{s}-wave scattering frequency. 

In Fig.\ref{fig:fig3} the chemical potential of the BEC (normalized to $ \hbar\omega_{R} $) has been plotted versus the pumping strength for three different values of the \textit{s}-wave scattering frequency. As is seen, below the threshold the chemical potential is zero in the absence of atomic collisions but it gets positive values in the presence of atomic interactions. on the other hand, above the threshold the chemical potential decreases by increasing the pumping strength and gets negative values.

Below the critical point the BEC behaves like a uniform system of bosons in the thermodynamic equilibrium in which all of them populate the single-particle state with the wave function $ 1/\sqrt{L} $. In such a situation, the interaction energy of a pair of particles is $ U_{s}/2L $ [the last term in Eq.(\ref{H1}) with $ \Psi=1/\sqrt{L} $]. Therefore, the energy of a state with $ N $ bosons all in the same state is given by multiplication of this quantity with the number of possible ways of making pairs of bosons, i.e., $ N(N-1)/2 $ \cite{Pethick}. In this approximation the energy of the system will be
\begin{equation}\label{E}
E=\frac{1}{2}N(N-1)\frac{U_{s}}{L}\simeq\frac{N^{2}U_{s}}{2L},
\end{equation}
where we have assumed $ N\gg 1 $. In the thermodynamic equilibrium the chemical potential of the system is $ \mu=\partial E/\partial N $. Therefore, below the critical point, i.e., for $ y<y_{c} $, the chemical potential of the system is
\begin{equation}\label{mu}
\mu=n_{a}U_{s}w^{2}=\frac{1}{2}\hbar\omega_{sw},
\end{equation}
where $ n_{a}=N/Lw^{2} $ is the density of the atoms. 

Our results obtained from the numerical self-consistent solutions to Eqs.(\ref{alpha}, \ref{beta}) completely agree with the above interpretation. As is seen from Fig.\ref{fig:fig3}, below the critical point, the chemical potential is zero in the absence of atomic collisions (solid line), while $ \mu=0.05\hbar\omega_{R} $ for $ \omega_{sw}=0.1\omega_{R} $ (dashed line) and $ \mu=0.1\hbar\omega_{R} $ for $ \omega_{sw}=0.2\omega_{R} $ (dotted line). 

In fact Eq.(\ref{beta}) is nothing more than the Gross-Piteavskii equation, i.e,
\begin{eqnarray}\label{GP}
\Big[-\frac{\hbar^2}{2m}\frac{d^2}{dx^2}+\hbar U_{0}|\alpha(t)|^2\cos^2(kx)+2\hbar\eta_{t}\alpha_{R}\cos(kx)\nonumber\\
+U_{s}|\psi(x)|^2\Big]\psi(x)=\mu\psi(x),\nonumber\\
\end{eqnarray}
where $ \psi(x)=\langle\Psi(x)\rangle $ is the macroscopic wave function of the condensate. In the two-mode model of the BEC the matter field can be written as follows
\begin{equation}
\Psi(x)=\psi(x)+\delta\Psi(x),
\end{equation}
where the macroscopic wave function and the quantum fluctuations are, respectively, given by
\begin{subequations}
\begin{eqnarray}
\psi(x)&=&\sqrt{\frac{N}{L}}\Big(\beta_{0}+\sqrt{2}\beta_{1}\cos(kx)\Big),\\
\delta\Psi(x)&=&\frac{1}{\sqrt{L}}\Big(\delta c_{0}+\sqrt{2} \delta c_{1} \cos(kx)\Big).
\end{eqnarray}
\end{subequations}

Below the critical point, where the mean value of the optical field inside the cavity is zero ($ \alpha=0 $) and all the atoms are in the condensate mode ($ \beta_{0}=1, \beta_{1}=0 $), the macroscopic wave functions is $\psi(x)= \sqrt{N/L} $ (super fluid phase). In this situation all the terms in the left hand side of Eq.(\ref{GP}) are zero except for the last one. Therefore $ \mu=U_{s}|\psi(x)|^2=U_{s}N/L $ which is just the result obtained in Eq.(\ref{mu}).

On the other hand, above the critical point where the mean value of the optical mode inside the cavity is no longer zero, the condensate mode of the BEC starts to be depleted due to the interaction with the optical field which leads to the decrease of the chemical potential. In this situation the cosine term is also present in the macroscopic wave function ($ \beta_{1}\neq 0 $) which means that the BEC is in the Mott-insulator phase.

\subsection{Stationary quantum fluctuations and the power law behavior}
In order to investigate the stationary properties of the system fluctuations one should consider the steady-state condition governed by Eq.(\ref{deltaZ}) which is reached when the system is stable. It occurs if and only if all the eigenvalues of the matrix $ A $ have negative real parts. These stability conditions can be obtained by using the Routh-Hurwitz criterion \cite{RH}. Due to the linearized dynamics of the fluctuations and since all noises are Gaussian the steady state is a zero-mean Gaussian state which is fully characterized by the $6\times6$ stationary correlation matrix (CM) $V$, with components $V_{ij}=\langle \delta Z_i(\infty)\delta Z_j(\infty)+\delta Z_j(\infty)\delta Z_i(\infty)\rangle/2 $. Using the QLEs, one can show that $ V $ fulfills the  Lyapunov equation
\begin{equation}\label{lyap}
AV+VA^T=-D,
\end{equation}
where
\begin{equation}\label{D}
 D=\mathrm{Diag}[\kappa,\kappa,\gamma_{c},\gamma_{c},\gamma_{c},\gamma_{c}],
\end{equation}
is the diffusion matrix. Here, we have also assumed that the mean number of thermal excitations of both the condensate and the Bogoliubov modes are zero which is reasonable for the BEC with temperature about nano Kelvin. Equation(\ref{lyap}) is linear in $V$ and can be straightforwardly solved. However, the explicit form of $ V $ is complicated and is not reported here.

\begin{figure}[ht]
\centering
\includegraphics[width=2.5in]{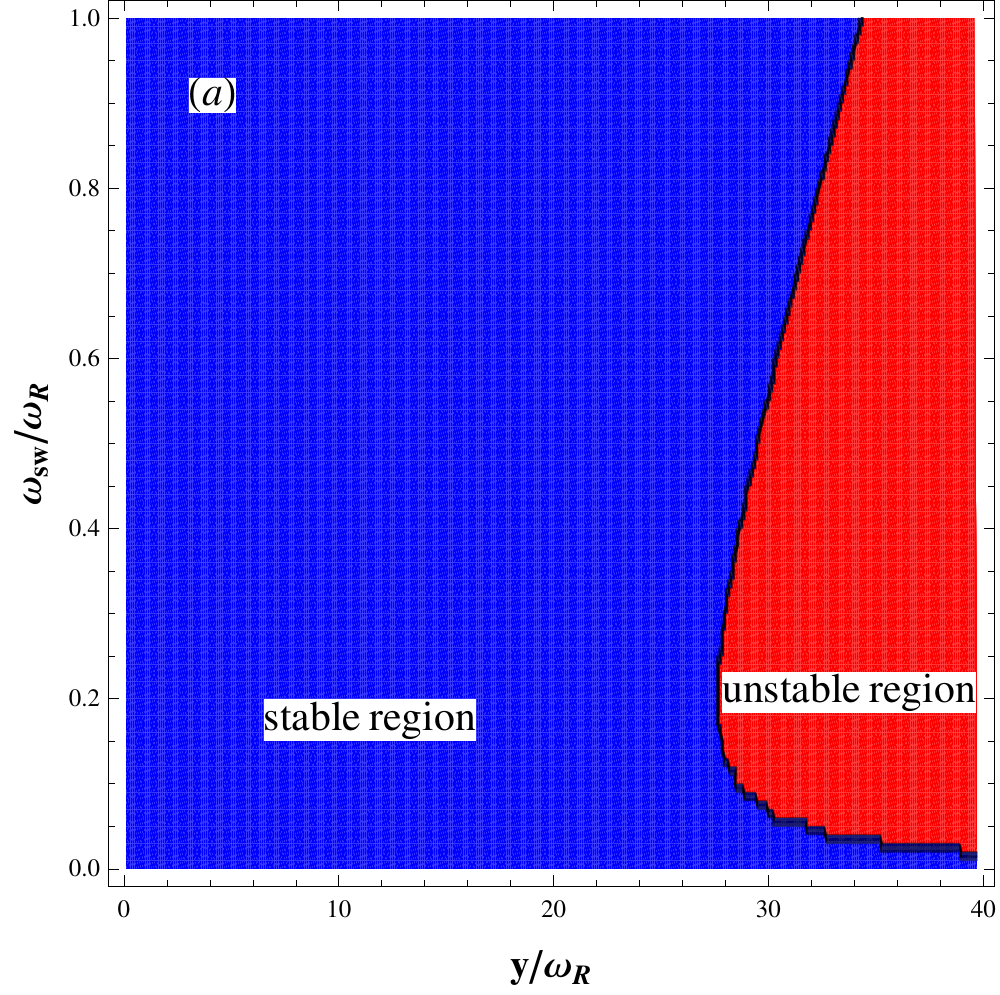}
\includegraphics[width=2.5in]{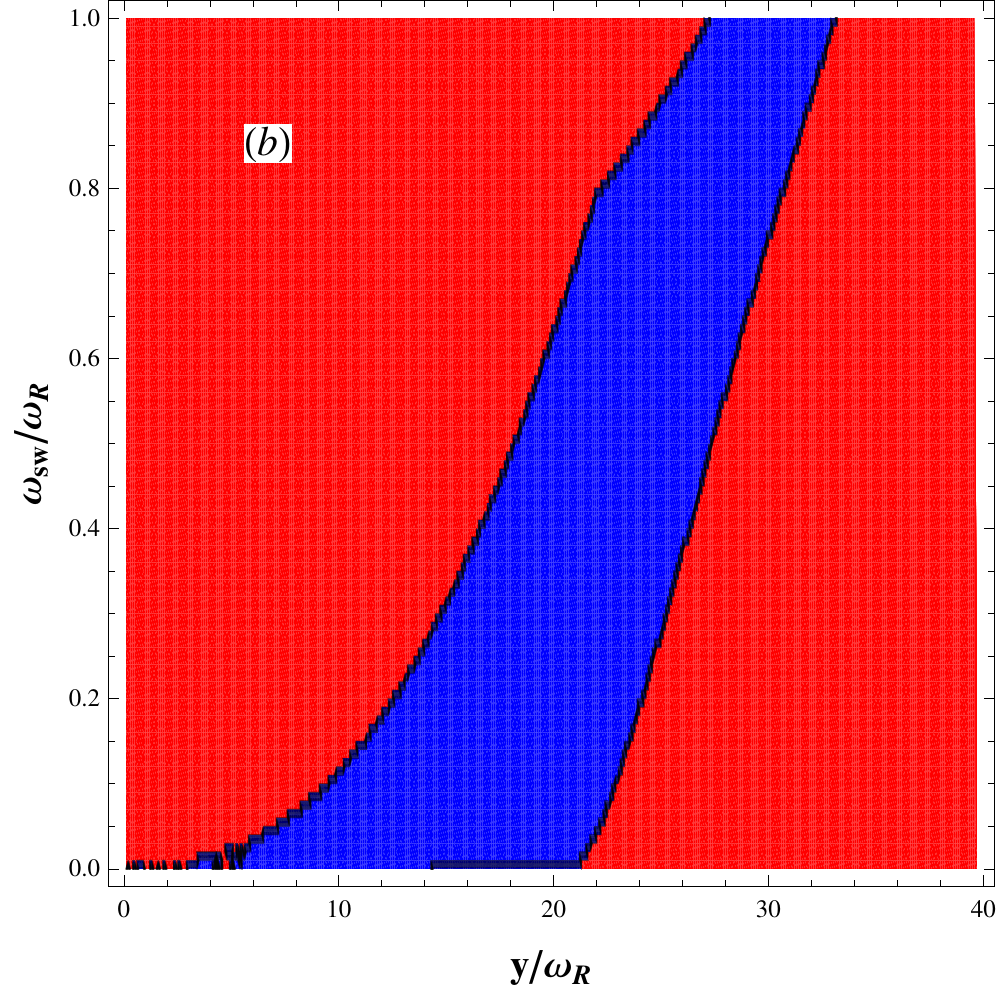}
\caption{
(Color online) The regions of dynamical stability (blue) and dynamical instability (red) versus the normalized pumping strength ($ y/\omega_{R} $) and the normalized \textit{s}-wave scattering frequency of atomic collisions ($ \omega_{sw}/\omega_{R} $) for two values of the BEC damping constant: (a) $ \gamma_{c}=0.001\kappa $ and (b) $ \gamma_{c}=0 $ The other parameters are the same as those of Fig.\ref{fig:fig2}.}
\label{fig:fig4}
\end{figure}

After calculating the mean-field values we can obtain the elements of the drift matrix $ A $ and calculate its eigenvalues. In Fig.\ref{fig:fig4} we have shown the dynamical stability regions (blue region)  and instability regions (red region) of the system in terms of the normalized pumping strength ($ y/\omega_{R} $) and the normalized \textit{s}-wave scattering frequency of atomic collisions ($ \omega_{sw}/\omega_{R} $) for two values of $ \gamma_{c}=0.001\kappa $ [Fig.\ref{fig:fig3}(a)] and $ \gamma_{c}=0 $ [Fig.\ref{fig:fig3}(b)]. The dynamical stable regions are those points of the space $ y-\omega_{sw} $ where all the eigenvalues of the drift matrix $ A $ have negative real parts and the dynamical unstable regions are those points where at least one of the eigenvalues has a positive real part. As is seen, when $ \gamma_{c}=0 $ the system is unstable even for low values of $ \omega_{sw} $ below the critical point. However, considering a nonzero value for $ \gamma_{c} $ even as small as $ 0.001\kappa $ preserves the stability of the system for wider domains of $ y $ and $ \omega_{sw} $.

\begin{figure}[ht]
\centering
\includegraphics[width=3in]{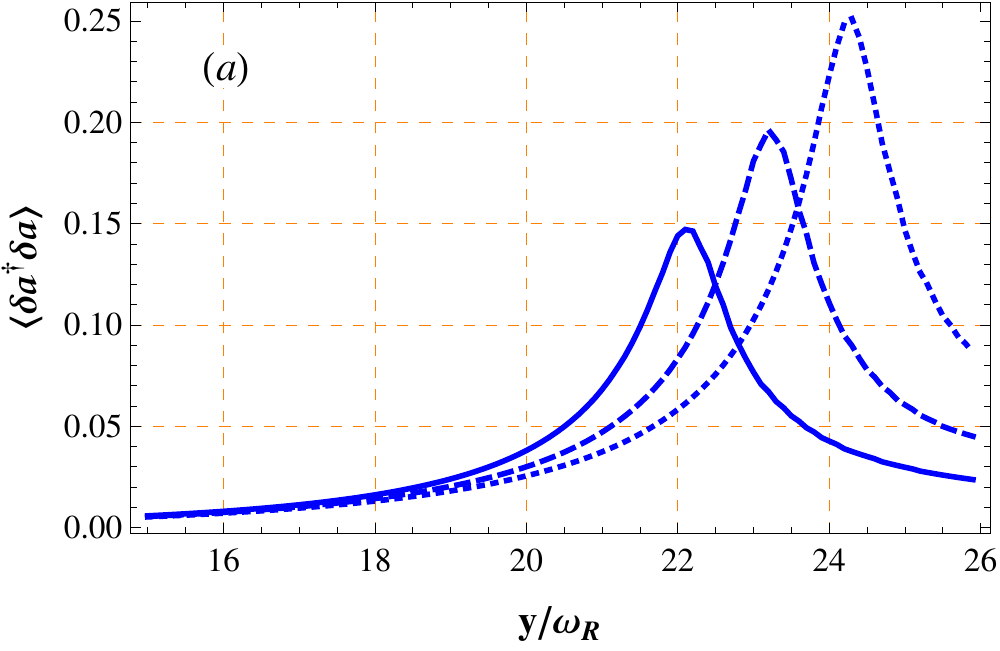}
\includegraphics[width=3in]{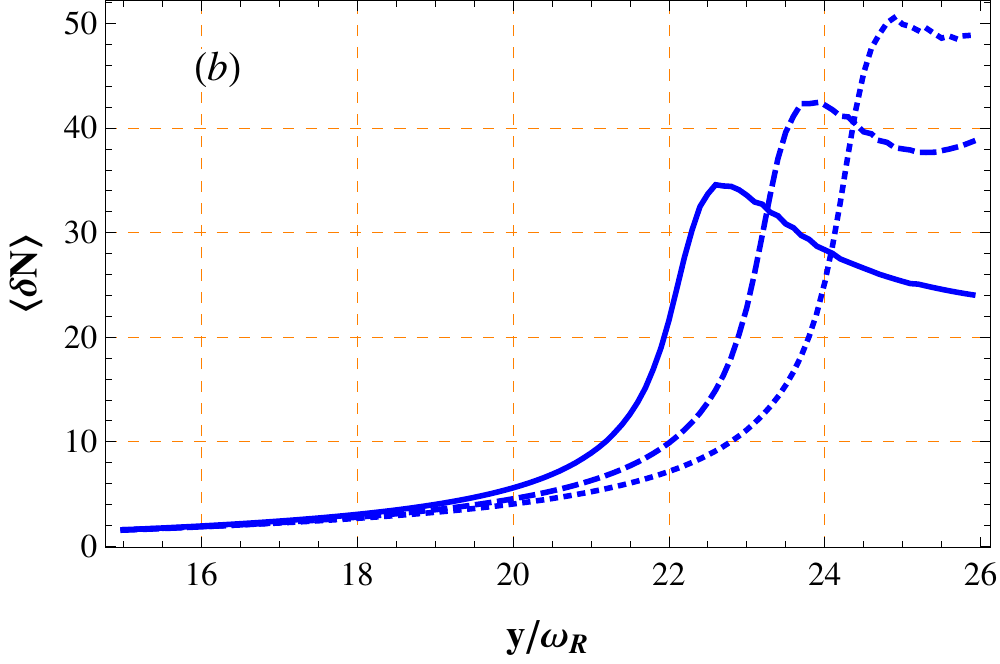}
\caption{
(Color online) (a) The effective number of incoherent photons of the cavity, and (b) the average of quantum fluctuations of the particle number operator of the BEC (condensate depletion) versus the normalized pumping strength $ y/\omega_{R} $ for three values of the \textit{s}-wave scattering frequency: $ \omega_{sw}=0 $ (solid lines), $ \omega_{sw}=0.1\omega_{R} $ (dashed lines), and $ \omega_{sw}=0.2\omega_{R} $ (dotted lines).The decay rate of the matter field is $ \gamma_{c}=0.001\kappa $ and the other parameters are the same as those of Fig.\ref{fig:fig2}. }
\label{fig:fig5}
\end{figure}

On the other hand, by solving the Lyapunov equation [Eq. (\ref{lyap})] we can obtain the correlation matrix $ V $ which gives us the second-order correlations of the fluctuations. In this way we can calculate the effective number of incoherent photons of the cavity,
\begin{equation}
\langle\delta a^{\dagger}\delta a\rangle=\frac{1}{2}(V_{11}+V_{22}-1),
\end{equation}
and the effective number of incoherent atoms in the condensate and the Bogoliubov modes,
\begin{subequations}
\begin{eqnarray}
\langle\delta c^{\dagger}_{0}\delta c_{0}\rangle=\frac{1}{2}(V_{33}+V_{44}-1),\\
\langle\delta c^{\dagger}_{1}\delta c_{1}\rangle=\frac{1}{2}(V_{55}+V_{66}-1).
\end{eqnarray}
\end{subequations}

The particle number operator, Eq.(\ref{N}), can be written as $ \hat{N}=N+\delta\hat{N} $ where
\begin{equation}\label{delta N}
\delta\hat{N}=\int\delta\Psi^{\dagger}(x)\delta\Psi(x)dx=\delta c^{\dagger}_{0}\delta c_{0}+\delta c^{\dagger}_{1}\delta c_{1},
\end{equation}
is the quantum fluctuations in the total particle number. The average value of this operator ($ \langle\delta\hat{N}\rangle $) can be defined as the depletion of the BEC, i.e., the total occupancy of motional modes other than the macroscopically populated BEC wave function \cite{szirmai PRL09}.

\begin{table}[ht]
\caption{The critical values and the critical exponents for the curves of Fig.\ref{fig:fig5}(b)}
\centering
\begin{tabular}{ccc}
\hline\hline
$ \omega_{sw}/\omega_{R} $ & $ y_{c}/\omega_{R} $ & $ \nu_{c} $ \\ [0.5ex]
\hline 
 0 & 22.12 & -0.466 \\ 

 0.1 & 23.27 & -0.592 \\ 

 0.2 & 24.28 & -0.617 \\ 
\hline 
\end{tabular}
\label{table:crit} 
\end{table}

In Fig.\ref{fig:fig5} we have plotted the effective number of incoherent photons of the cavity and the average of quantum fluctuations of the particle number operator of the BEC (condensate depletion) versus the normalized pumping strength $ y/\omega_{R} $ for three different values of atom-atom interaction. The results have been shown in that range of the pumping strength where the system is stable for the three different values of the \textit{s}-wave scattering frequency as depicted in Fig.\ref{fig:fig4}(a) where the decay rate of the matter field is $ \gamma_{c}=0.001\kappa $ .

As is seen, below the critical point the fluctuations of photons and atoms are zero. However, as the transverse pumping strength reaches near the critical value $ y_{c} $ a sharp pick appears. Increasing the \textit{s}-wave scattering frequency from zero (solid lines) to $ 0.1\omega_{R} $ (dashed lines) and $ 0.2\omega_{R} $ (dotted lines) shifts the pick (the phase transition threshold) to higher values of the pumping strength just like what was happened for the mean value solutions [Figs.\ref{fig:fig2}, \ref{fig:fig3}]. Besides, the greater the strength of atom-atom interaction is, the higher the pick of fluctuations.

Near the critical point and in the limit $ \gamma_{c}\rightarrow 0 $,the fluctuations in the number of photons and atoms get divergent (the heights of the picks become so much large). This is one of the most important characteristics of a second-order phase transition phenomenon where the correlation functions of the system exhibit a power law, i.e., $ (y-y_{c})^{\nu_{c}} $, near the critical point where $ \nu_{c}<0 $ is the critical exponent. Based on our numerical results, choosing $ \gamma_{c}=0.001\kappa  $ not only provides the stability conditions of the system [Fig.\ref{fig:fig4}(a)] but also is small enough to show us the scaling law behavior of the system near the critical point. The physical origin of this damping constant is the coupling of the two modes of the BEC to the other modes ($ C_{n,m} $) due to the atomic collisions or through the harmonic trapping potential of the condensate \cite{k zhang 10}.

\begin{figure}[ht]
\centering
\includegraphics[width=3in]{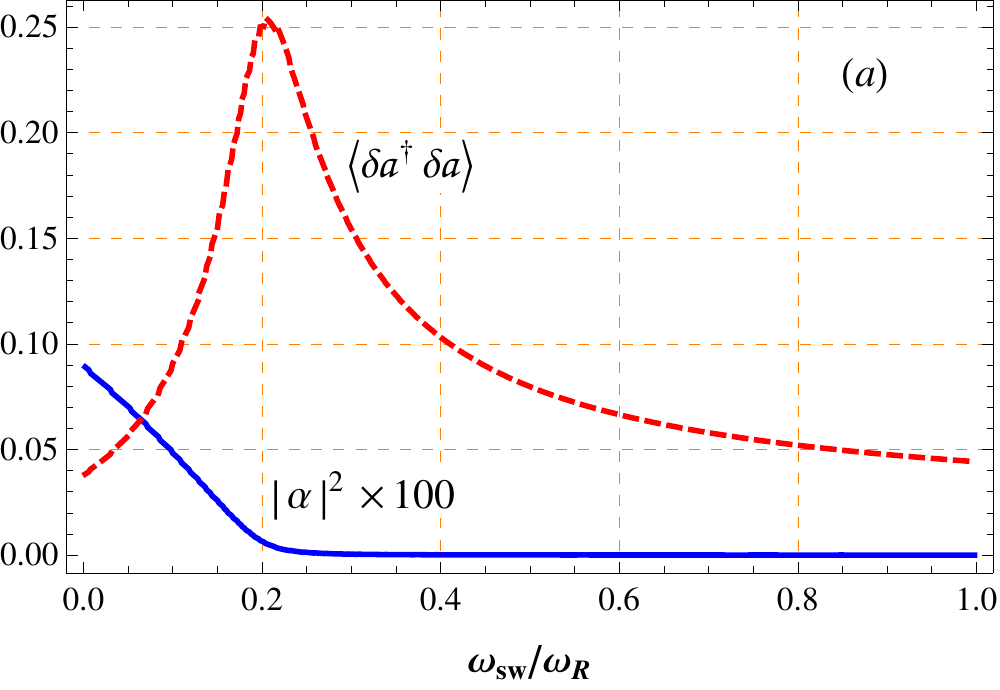}
\includegraphics[width=3in]{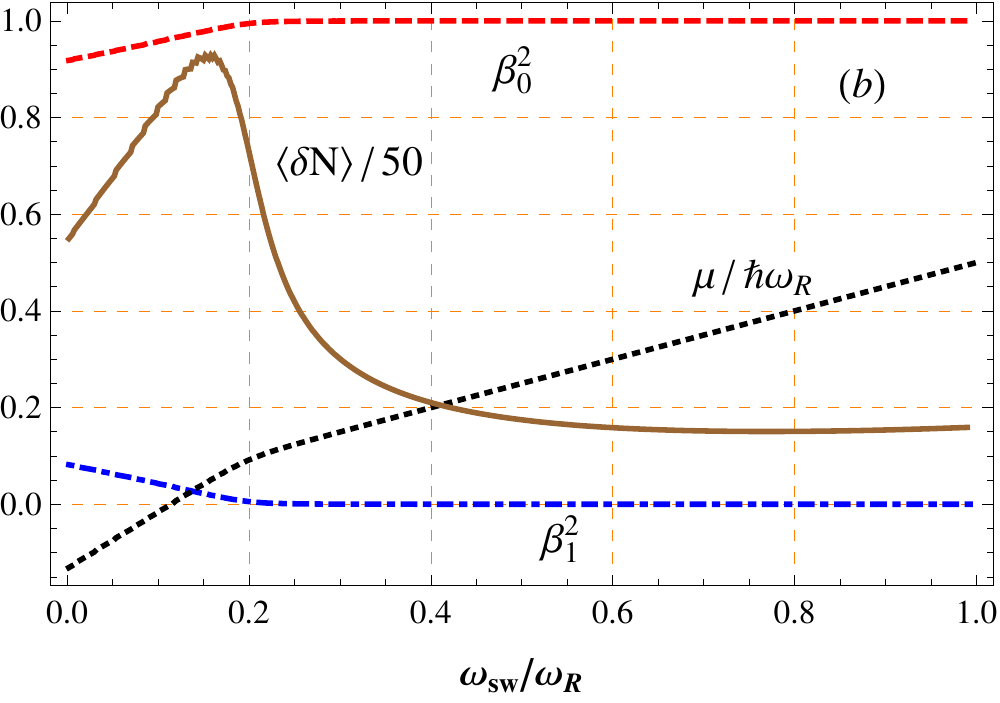}
\caption{
(Color online) (a) The mean photon number of the cavity normalized to $ \sqrt{N} $ magnified by 100 (blue solid line) and the effective number of incoherent photons of the cavity (red dashed line) versus the \textit{s}-wave scattering frequency normalized to $ \omega_{R} $. (b) The mean value fraction of atoms in the condensate mode (red dashed line) and in the Bogoliubov mode (blue dashed-dotted line), the chemical potential normalized to $ \hbar\omega_{R} $ (black dotted line), and the condensate depletion divided by 50 (brown solid line) versus the \textit{s}-wave scattering frequency normalized to $ \omega_{R} $. The strength of the transverse pumping has been set to $ y=24.28\omega_{R} $, the decay rate of the matter field is $ \gamma_{c}=0.001\kappa $ and the other parameters are the same as those of Fig.\ref{fig:fig2}. }
\label{fig:fig6}
\end{figure}

In Table \ref{table:crit}, we have listed the critical values of transverse pumping strength ($ y_{c} $) and the critical exponents ($ \nu_{c} $) for the power law behavior of the condensate depletion $ \langle\delta\hat{N}\rangle $ near the critical point for three different values of the \textit{s}-wave scattering frequency. For this purpose we have fitted functions like $ r (y-y_{c})^{\nu_{c}} $ to the three curves of Fig.\ref{fig:fig5}(b) when $ y $ gets near to $ y_{c} $ from the left hand side and have obtained the corresponding critical exponents. As is seen from this Table, the increase of $ \omega_{sw} $ not only shifts $ y_{c} $ to the larger values but also leads to the increase of the absolute value of the critical exponents.

The last point we are going to investigate is the case that has been shown in Fig.\ref{fig:fig6} where we have set the strength of the transverse pumping to a fixed value, say $ y=24.28\omega_{R} $, and have examined the variation of the mean fields and quantum fluctuations of the system in terms of the variation of the \textit{s}-wave scattering frequency. As is seen, here the \textit{s}-wave scattering frequency acts as the control parameter for the quantum phase transition. There is a critical value $ \omega_{sw}^{(c)}=0.2\omega_{R} $ where below it the system is in the superradiant phase and above that value the system goes to the normal phase. It is in accordance with the results given in the last line of Table \ref{table:crit}.

Besides, near the critical point where $ \omega_{sw}\rightarrow \omega_{sw}^{(c)} $ the quantum fluctuations in photons [red dashed line in Fig.\ref{fig:fig6}(a)] and also in the total number of atoms (condensate depletion) [Brown solid line in Fig.\ref{fig:fig6}(b)] get maximum. In this way for each value of the transverse pumping strength there is a critical value for the \textit{s}-wave scattering frequency where below it the system is in the superradiant phase and above that value the system goes to the normal phase.

\section{Conclusion}\label{secCon}
In this paper, we have studied the effect of atomic collisions on the phase transition form the normal to the superradiant phase in a one-dimensional BEC inside an optical cavity which is pumped from the transverse side. In the first part of the paper, we have investigated explicitly which modes of the BEC are excited when the atoms are driven from the transverse side of the cavity and have shown how the BEC can be described approximately through a two-mode model. 

In the second part of the paper, we have made use of the two-mode model of the BEC and showed that increasing the atom-atom interaction strength causes the threshold of the phase transition (the critical transverse pumping strength) to be shifted to the larger values. Besides, we have studied the power law behavior of the system near the critical points and showed that how the \textit{s}-wave scattering frequency of atomic collisions affect the critical exponents of the quantum fluctuations in the total number of atoms. Finally, it has been shown that by fixing the strength of the transverse pumping on a specific value, the quantum phase transition can be observed by variation of the \textit{s}-wave scattering frequency, i.e., when $ \omega_{sw} $ passes by a critical value the phase of the system changes from the superradiant to the normal phase.

\section*{Acknowledgement}
The authors wish to thank The Office of Graduate Studies of The University of Isfahan for its support.

\bibliographystyle{apsrev4-1}

\end{document}